# Asymmetrical plasmon distribution in hybrid AuAg hollow/solid coded nanotubes

*Aziz Genç,[a,b,*] Javier Patarroyo,[a] Jordi Sancho-Parramon,[c] Raul Arenal,[d,e] Neus G. Bastús,[a] Victor Puntes,[a,f,g] Jordi Arbiol[a,g,*]*

[a] Catalan Institute of Nanoscience and Nanotechnology (ICN2), CSIC and BIST, Campus UAB, Bellaterra, 08193 Barcelona, Catalonia, Spain.

[b] Materials Science and Engineering Department, Izmir Institute of Technology, 35430, İzmir, Turkey.

[c] Rudjer Boskovic Institute, Zagreb, Croatia.

[d] ARAID Foundation, 50018 Zaragoza, Aragon, Spain.

[e] Laboratorio de Microscopias Avanzadas (LMA), Instituto de Nanociencia y Materiales de Aragon (INMA), Universidad de Zaragoza, 50018 Zaragoza, Spain.

[f] Vall d'Hebron Institut de Recerca (VHIR), 08035, Barcelona, Catalonia, Spain.

[g] ICREA, Pg. Lluís Companys 23, 08010 Barcelona, Catalonia, Spain.

KEYWORDS Plasmon coded, nanotubes, nanowires, asymmetrical distribution; metal nanotubes, electron energy-loss spectroscopy, AuAg, localized surface plasmon resonances, boundary element method

ABSTRACT

Morphological control at the nanoscale paves the way to fabricate nanostructures with desired plasmonic properties. We report the nanoengineering of plasmon resonances in 1D hollow nanostructures of two different AuAg nanotubes; completely hollow nanotubes and hybrid nanotubes comprising solid Ag and hollow AuAg segments. Spatially resolved plasmon mapping by electron energy loss spectroscopy (EELS) revealed the presence of high order resonator-like modes and localized surface plasmon resonance (LSPR) modes in both nanotubes. Experimental findings are accurately correlated with boundary element method (BEM) simulations, where both experiments and simulations revealed that the plasmon resonances are intensely present inside the nanotubes due to plasmon hybridization. Based on the experimental and simulated results reported, we show that the novel hybrid AuAg nanotubes possess two significant coexisting features: (i) LSPRs have been generated distinctively from the hollow and solid parts of the hybrid AuAg nanotube, which opens the way to control a broad range of plasmon resonances with one single nanostructure and (ii) the periodicity of the high-order modes are disrupted due to the plasmon hybridization by the interaction of solid and hollow parts, resulting in an asymmetrical plasmon distribution in 1D nanostructures. The asymmetry could be modulated/engineered leading to control coded plasmonic nanotubes.

INTRODUCTION

Surface plasmon resonances are collective oscillations of conduction electrons in a material excited by an electromagnetic wave [1]. Such a unique property has enabled the usage of plasmonic nanostructures as building blocks for nanooptics and various novel applications including, but not limited to, sensor devices [2, 3], surface-enhanced Raman spectroscopy (SERS) [4, 5], biomedicine [6], photovoltaics [7, 8], thanks to their ability of localization of light at the nanoscale, far beyond the diffraction limit of visible electromagnetic waves in dielectric media [9, 10].

Localized surface plasmon resonances have the ability to direct and enhance radiative emission (and vice versa) in transmission mode and they can convert the propagating freespace EM wave to highly confined and strongly enhanced electric fields [11, 12, 13, 14, 15, 16]. Plasmonic nanostructures can be termed as nanoantennas as their EM modulation mechanism is similar to the radio antennas [11, 12]. Gold and silver nanostructures are promising nanoantennas with their good metallic properties [14]. Thanks to their ability to convert light into highly localized fields, plasmonic nanoantennas have been used to enhance the performance of various photoactive devices such as solar cells, photodetectors and biosensors [7, 17, 18]. They are also widely used in nanophotonic circuits as they are able to modify the amount and direction of the emitted electromagnetic energy [19, 20, 21].

Electron energy-loss spectroscopy (EELS) has been widely used to obtain plasmonic properties of various nanostructures [22, 23]. Among the different metal nanostructures analyzed via EELS, the plasmonic properties of Ag nanorods/nanowires are intensively studied [24, 25, 26, 27, 28, 29]. However, there is not much information about the plasmonic properties of 1D hollow metal nanostructures in the literature. Hollow metal nanostructures, in general, exhibit enhanced plasmonic properties [30, 31, 32, 33] due to a mechanism called plasmon hybridization between solid parts and voids of the nanostructures [34]. S. Yazdi et al. [35] reported the controllable, reversible, and dynamic tuning of plasmon resonances in partially hollow AgAu nanorods via EELS, which showed the extent of the ability to control plasmonic properties in hollow nanostructures [35]. To further investigate this phenomenon by comparing the plasmonic properties of hollow and partially hollow nanostructures, the present study reports the observation of plasmon resonances in 1D hollow nanostructures of two different AuAg nanotubes; completely hollow nanotubes and hybrid nanotubes comprising the sequential formation of solid Ag parts and hollow AuAg parts. We believe that these novel AuAg nanotubes can be good alternatives to the above-mentioned applications as they possess controllable and enhanced plasmonic properties covering the ranges of ultraviolet, visible and near-infrared in a single nanostructure.

METHODS

**Synthesis of 1D AuAg nanostructures**

The syntheses of AuAg nanotubes were produced by the galvanic replacement reaction of Ag nanowires used as templates following a similar methodology to that previously described in [36, 37]. In the present case, several micron-long, penta-twinned Ag nanowires with a diameter of about 80 nm were synthesized via a solution-phase approach as reported by Sun et al. [38] and used as templates. In a typical synthesis of completely hollow AuAg nanotubes, 0.25 mL of Ag nanowires (310 ppm $Ag^+$, 2.9 mM by ICP-MS) were dispersed in a solution containing 2 mL of milli-Q water, 1 mL of CTAB (14 mM), and 0.1 mL of AA (0.1 mM). Then, 0.5 mL of $HAuCl_4$ (1mM) was added through a syringe pump at 25 $\mu$L/min rate under stirring. After adding the HAuCl$_4$ solution, the reaction was stirred at room temperature for about 30 min until the UV-vis spectra of the solution remained unaltered. Next, the sample was then centrifuged at 5000 g and washed with milli-Q water twice; finally the pellet was re-suspended in 1 mL of milli-Q water for further characterization. For the synthesis of hybrid AuAg nanotubes, the procedure was the same except the amount of HAuCl$_4$ (1mM), which was decreased.

Solutions containing the 1D AuAg nanostructures were ultrasonicated for about 15 minutes and deposited on 15 nm thick Si$_3$N$_4$ membrane grids for STEM-EELS investigations. A hydrogen plasma cleaning using a Plasma Etch$^{TM}$ plasma cleaner is applied prior to EELS analyses to eliminate the organic residues present from the synthesis procedure. It should be mentioned here that the applied ultrasonication procedure may cause the breakage of the longest nanotubes.

**EELS acquisition and data processing**

Electron energy-loss spectroscopy (EELS) in a scanning transmission electron microscope (STEM) equipped with a monochromator is an ideal technique for studying plasmonic responses of nanostructures, having access to high spatial and energy resolutions [39]. A probe-corrected FEI™ Titan 60–300 STEM equipment was used for the EELS analysis, which was operated at 80 kV. The microscope is equipped with a high-brightness X-FEG gun, a Wien filter monochromator and a Gatan™ Tridiem 866 ERS energy filter. A collection angle of 32 mrad and a dispersion of 0.01 eV per channel was used during the acquisition. Typical energy resolutions (full-width at half-maximum of the ZLP) of the measurements were better than 150 meV. EEL spectra were acquired using the spectrum imaging (SI) method [40,41] in which a sub-nanometer electron probe was scanned over the area of interest with a constant displacement of 4 – 6 nm.

EELS data has been processed by using a used a spectral un-mixing (SU) based routine of vertex component analysis (VCA) [42, 43, 44, 30], which is implemented in the HyperSpy [45] multi-dimentional data analysis toolbox.

**Simulations**

Throughout the paper, we have used the boundary element method (BEM) [46, 47] simulations to understand the effects of shape, composition and environmental (substrate) on the plasmonic properties of hollow 1D AuAg nanostructures. All the BEM simulations were done using the MNPBEM Matlab$^{TM}$ toolbox developed by Hohenester [48]. The optical constants of the bulk metals have been taken from the Johnson & Christy [49] and modified according to Ref. [50] for AuAg alloys. The size effects on the dielectric properties, assuming an increase of the damping constant in the Drude model with reduction of the particle size due to the electron confinement effects, were also taken into account during the BEM simulations [51].

## RESULTS AND DISCUSSION

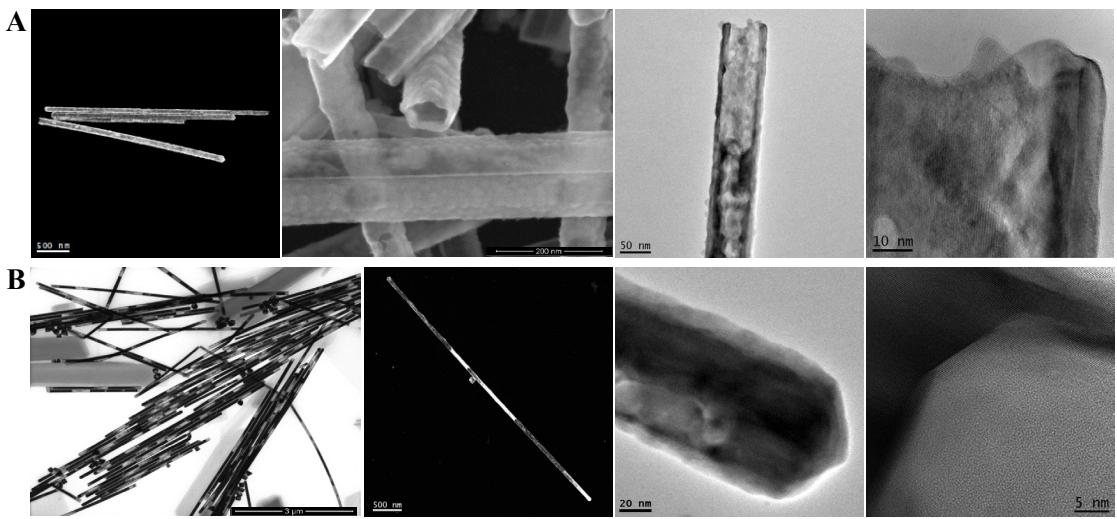

**Figure 1. Microstructure of AuAg nanotubes. A**. HAADF-STEM, SEM, TEM and HRTEM micrographs of hollow nanotubes, respectively. **B**. BF-STEM, HAADF-STEM, TEM and HRTEM micrographs of the hybrid nanotubes, respectively.

General microstructural features of the AuAg prepared nanotubes are presented in Fig. 1. Upper row (Fig. 1A) shows the high-angle annular dark field (HAADF) STEM, SEM, TEM and high-resolution TEM (HRTEM) micrographs obtained from the completely hollow AuAg nanotubes. Interestingly, images reveal that (i) nanotubes with lengths up to several micrometers preserving the penta-twinned structure of the Ag nanowires used as sacrificial templates and (ii) highly crystalline nanotubes having a wall thickness of ~10 nm and occasional pores along the walls. Bright-field (BF) STEM, HAADF-STEM, TEM and HRTEM micrographs obtained from the hybrid AuAg nanotubes are shown in the lower row (Fig. 1B), which clearly reveals that these nanotubes constitute sequential formation of hollow parts within the solid Ag nanowire templates.and solid parts. Hollow parts are composed of AuAg external walls formed after the galvanic replacement of Ag with Au and the solid parts are pure Ag covered by an AuAg thin shell (see STEM energy dispersive X-ray spectroscopy (EDX) results presented in Fig. S1-S2). It is also revealed in Fig. 1B that hybrid AuAg nanotubes also preserve the penta-twinned structure and are highly crystalline. As mentioned in the experimental procedure section, these 1D structures are synthesized by following the procedure reported in [36]. Very recently, Canepa et al. [52] reported anisotropic galvanic replacement reactions in Ag nanowires, synthesizing similar hybrid nanotubes as the present study. Here, we present a detailed investigation about the nanoscale distribution of plasmon resonances and plasmon mode interactions of such 1D nanostructures.

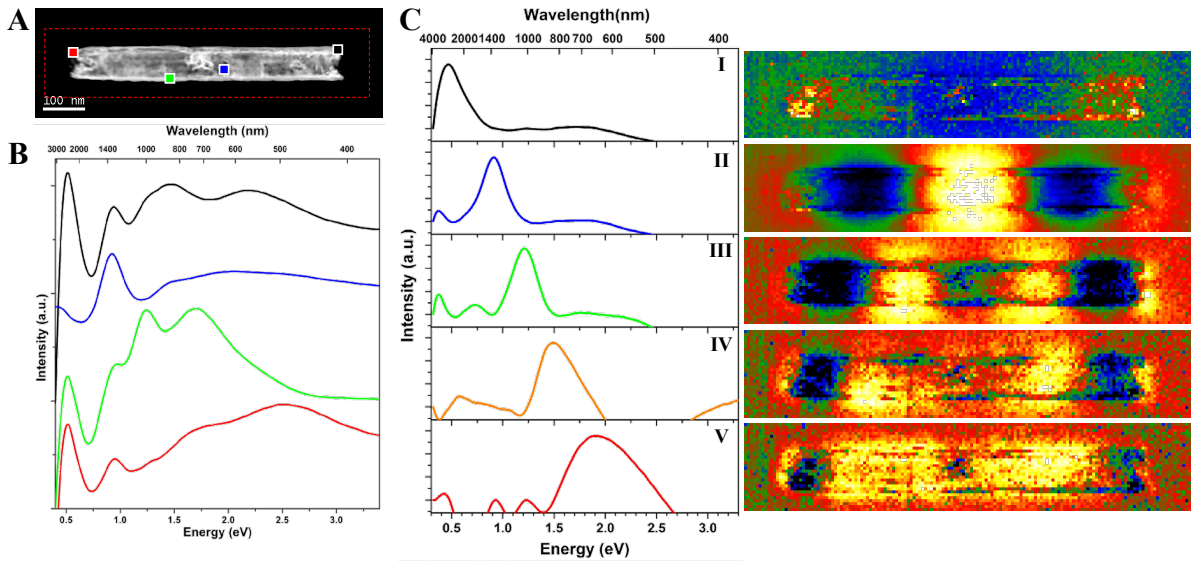

**Figure 2. Plasmonic properties of a completely hollow AuAg nanotube. A**. HAADF-STEM micrograph of the AuAg nanotube, which is 84 nm in diameter and 665 nm in length. The area of the EELS SI is indicated with a red rectangle. B. Background subtracted selected area EEL spectra of different locations marked in A. C. Spectra and corresponding abundance maps of 5 plasmonic components obtained by VCA processing.

As seen in the above-presented micrographs, both completely hollow and hybrid AuAg nanotubes have lengths of several microns, which is quite impractical for nanoscale EELS mapping due to the long acquisition time and stability (drift) issues. Therefore, we have tried to choose shorter nanotubes yet keep the general features. Fig. 2 shows the plasmonic properties of a completely hollow AuAg nanotube, which is 84 nm in diameter (wall thickness: ~10 nm) and 665 nm in length (Fig. 2A). Spatially resolved plasmonic properties of the nanotube are studied by obtaining an EELS spectral imaging (SI) over the area indicated with a red rectangle in Fig. 2A. Fig. 2B shows the background-subtracted selected area EEL spectra of different locations depicted in Fig. 2A, revealing various plasmon peaks at ~0.5 eV, ~0.9 eV, ~1.2 eV, ~1.45 eV, ~1.7 eV, ~2.2 eV and ~2.54 eV. It should be stressed here that one needs to take extra precautions while applying background subtraction routine by using the power law [53] as the energy of the plasmon peaks are close to the zero-loss peak.

Obtained EELS data is processed by using a spectral un-mixing routine based on the vertex component analysis (VCA) algorithm. Fig. 2C shows spectra of 5 different plasmon components and their corresponding abundance maps obtained by applying VCA analysis to the EELS data of the completely hollow AuAg nanotube. Plasmon components obtained by VCA reveal the presence of Fabry-Perot resonator resonances [54] and LSPR for the AuAg nanotube (note that the colors of the spectra in the Fig. 2C do not represent the colored regions marked in Fig. 2A). Fabry-Perot resonator modes in such quasi-1D metallic structures (or nanoantennas), consist of propagation as well as reflection of plasmons [54]. First, second and third-order modes (components I, II and III) are located at ~0.5 eV, ~0.9 eV and ~1.2 eV, respectively. It should be emphasized here that Fabry-Perot resonances seem to be most intense inside the nanotube, which is clearly revealed in the abundance maps of components II (second-order mode) and III (third-order mode). By looking at the abundance map of component IV, which is located at ~1.5 eV, one can suggest that this mode is more like a fourth-order mode than a LSPR mode. A wide peak located between 1.4 eV and 2.7 eV with a maximum at ~1.9 eV (component V) is associated with a LSPR mode, and its corresponding abundance map shows its distribution throughout the nanotube. Such plasmon distribution maps with nanoscale resolution are obtained for the first time for metal. The finding reported here are quite similar to our previous report on the AuAg nanoboxes [30], where the distribution of highly intense plasmon resonances in and around the nanostructures can be clearly observed, and the reason for their enhanced plasmonic properties (increased intensity and further reach). The fact that hollow 1D nanostructures have highly intense Fabry-Perot and LSPR modes both at the inner and outer parts of the nanotube suggests that they can be good alternatives in different applications of plasmonic nanoantennas [11, 12, 14].

We continue with the simulations of plasmonic properties in 1D metal nanostructures. Although we did not conduct EELS experiments on Ag nanowires as their properties are widely studied in the literature, we simulate the plasmonic properties of an Ag nanowire having same size as the hollow AuAg nanotube as a reference. By using this reference, we have also studied the differences between the solid and hollow Ag nanostructures and the effects of the substrate (i.e. 15 nm thick $SiN_x$ TEM grids) presence, all of which are detailly discussed in the Supporting Information section (Fig. S3-S5).

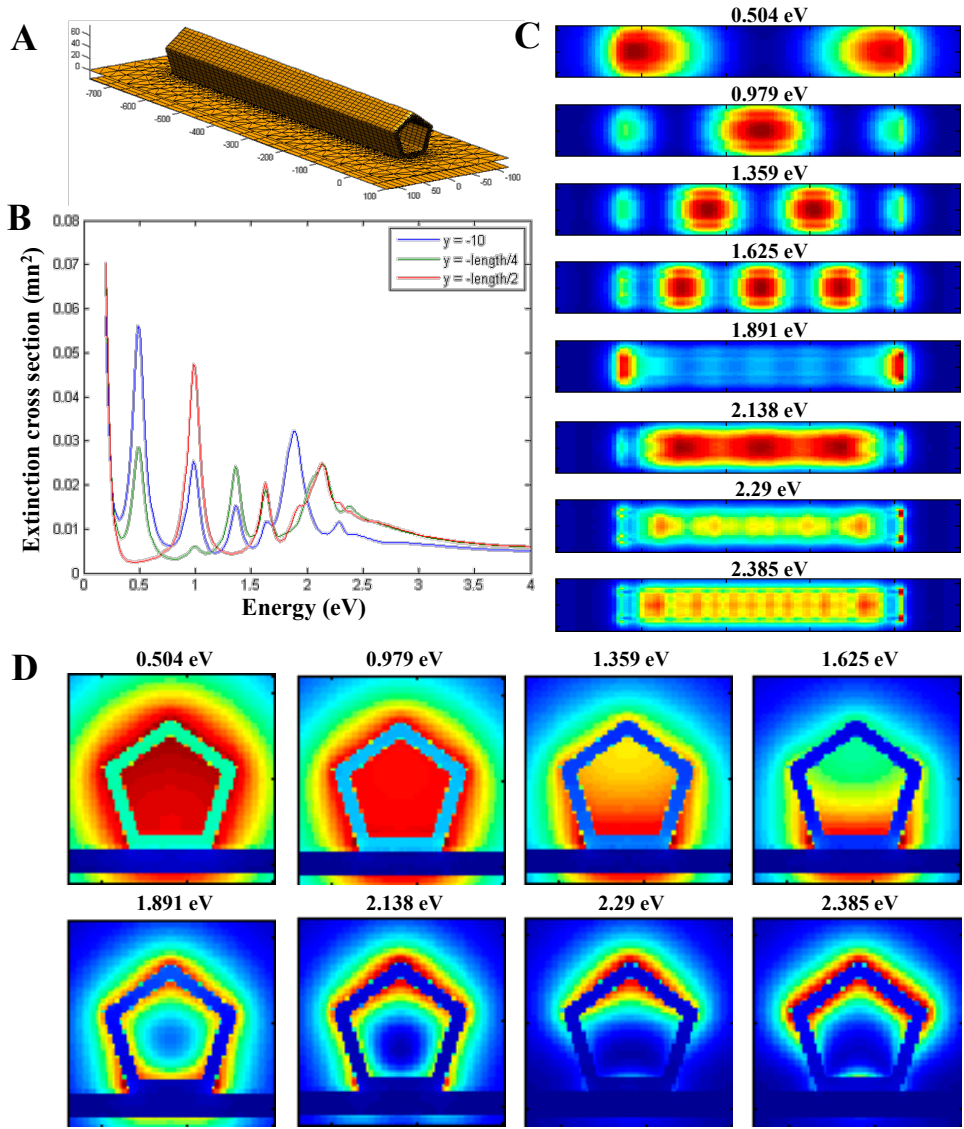

**Figure 3. BEM simulations of the completely hollow AuAg nanotube. A.** Structural model of the BEM simulated AuAg nanotube with a length of 665 nm and a diameter of 84 nm, with 10 nm thick walls, standing on a 15 nm thick $Si_3N_4$ substrate. **B.** Simulated local EEL spectra obtained at the tip (in blue), at a quarter of the length (at ∼ 166 nm, in green) and at the center (at 332.5 nm, in red) of the AuAg nanotube. **C.** BEM simulated plasmon maps of 8 different modes, located at 0.504 eV, 0.979 eV, 1.359 eV, 1.625 eV, 1.891 eV, 2.139 eV, 2.29 eV and 2.385 eV, from the in-plane view. **D.** BEM simulated plasmon maps of the same modes from the cross-sectional view.

In Fig. 3, BEM simulation results of the AuAg nanotube with a composition of 60 at.% Au and 40 at.% Ag with 10 nm thick continuous walls are presented. The simulated AuAg nanotube has the same dimensions as the experimentally studied one (Fig. 2). Several assumptions had to be made during the simulations: (i) the distribution of Au and Ag throughout the nanotube is considered as homogeneous and (ii) we discarded the possibility of pores around the walls . Fig. 3A shows the structural model of the AuAg nanotube standing on a 15 nm thick $Si_3N_4$ substrate. Fig. 3B shows the BEM simulated EELS spectra obtained at the tip (near the edge, in blue), at one-quarter of the nanotube length (at ∼166 nm, in green) and the center (at 332.5 nm, in red), revealing the presence of several peaks. It is worth noting that there is almost no visible peak at energies higher than 2.5 eV, unlike in the EEL spectra obtained from pure Ag (Fig. S4-S7). The plasmon peaks at high energies are diminished due to a mechanism called plasmon damping [55], in which the overlapping of the onset of the interband transitions with the LSPRs of Au causes a decrease in LSPR intensity at higher energies [55, 56]. Another feature to note is that the intensity of the EELS signal below 0.5 eV increases towards lower energy, which might suggest a possible presence of a dark plasmonic breathing-like mode at this energy range, yet it is hard to tell by just looking at these simulated EEL spectra. BEM simulated plasmon maps of 8 different modes are presented for the AuAg nanotube (Fig. 3C). Fabry-Perot resonator modes up to fourth order are revealed, which shows that these modes are excited most intensely inside the nanotube. These maps obtained by BEM simulations are quite similar to the experimentally observed abundance maps of VCA (Fig. 2C), where it is shown that Fabry-Perot modes are highly intense inside the nanotube. The LSPR mode located at 1.891 eV is confined at the tips of the nanotube and the LSPR mode located at 2.138 eV seems to be present all over the AuAg nanotube. Moreover, plasmon maps of two other LSPRs of different polar modes located at 2.29 eV and 2.385 eV are shown in this figure. As discussed in the Supporting Information part and reported in the literature [57, 58, 59] presence of a substrate splits the LSPR modes into proximal and distal modes. Fig. 3D shows the BEM simulated plasmon maps of the AuAg nanotube obtained by the beam incident on the pentagonal cross-section to better understand the proximal and distal modes. As seen in this figure, cross-sectional plasmon maps of the 8 different modes (whose planar views are presented in Fig. 3C) are shown, where the 15 nm thick substrate is clearly visible. First and second-order Fabry-Perot modes have a homogeneous distribution all around, both at the inner and outer parts of the AuAg nanotube. However, third and fourth-order Fabry-Perot modes have a highly intense distribution confined to the lower parts (which are in contact with the substrate) of the nanotube, resembling proximal modes. The first LSPR mode located at 1.891 eV is a proximal mode, with a more or less homogeneous distribution of plasmon resonances along the inner and outer parts of the nanotube wall. The other three LSPR modes located at 2.138 eV, 2.29 eV and 2.385 eV can be clearly identified as distal modes from these cross-sectional plasmon maps. It is shown that the mode located at 2.29 eV is confined at the distal corner with some contributions from the upper edges, whereas the mode located at 2.385 eV is more like an edge LSPR mode generated from the distal edges. BEM simulated 3D maps of these modes are presented in the Fig. S6.

Fig. 4A shows a 1.24 $\mu$m long hybrid AuAg nanotube with a diameter of 89 nm, where EELS SI is obtained over the area indicated with a red rectangle. Fig. 4B shows the background-subtracted selected area EEL spectra of different locations indicated in Fig. 4C, which is the EELS SI obtained from the region marked with a red rectangle in Fig. 4A. As seen in the local EEL spectra, this hybrid AuAg nanotube contains multiple plasmon resonances with energies from ~ 0.6 eV to 3.8 eV. Unlike the above presented completely hollow nanotube, the hybrid nanotube contains plasmon resonances generated with higher energies up to bulk plasmon resonance of Ag located at ~3.8 eV (shown in red). It is also worth noting that the EEL spectrum obtained from the area indicated with a purple square reveals the presence of a LSPR mode of Ag located at ~3.36 eV. By comparing the EEL spectra obtained from the hollow and solid parts of the hybrid nanotube (shown in pink and red, respectively), one can see that plasmon resonance of these regions are quite distinctive from one another.

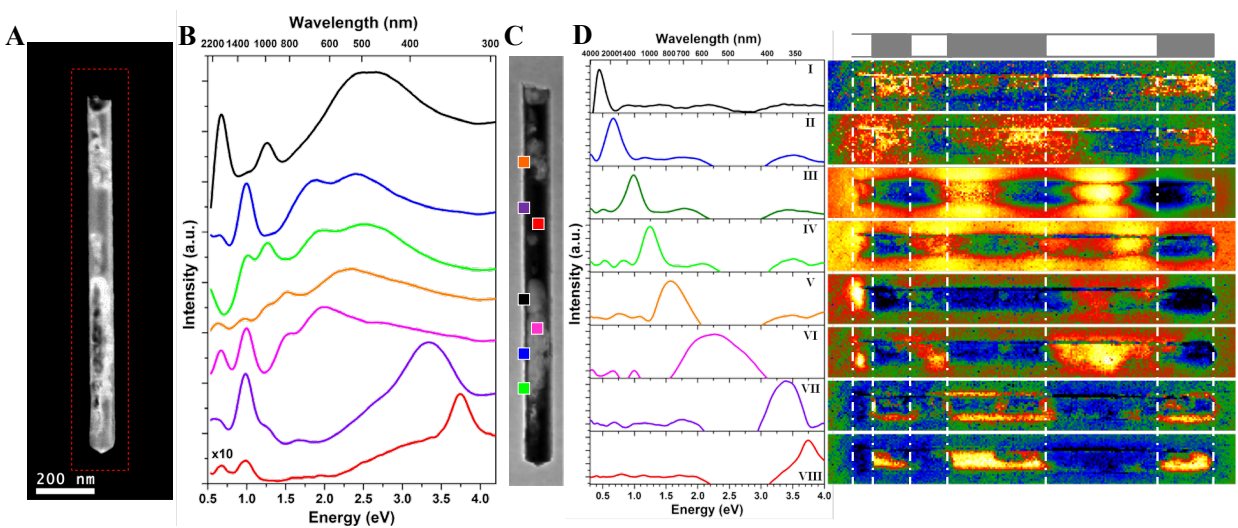

**Figure 4. Plasmonic properties of a hybrid AuAg nanotube. A.** HAADF-STEM micrograph of hybrid AuAg nanotube, which is 89 nm in diameter and 1.24 µm in length. The area of the EELS SI is indicated with a red rectangle. **B.** Background subtracted selected area EEL spectra of different locations marked in **C**, which is the EELS SI taken from the red rectangle in A. **D.** Spectra and corresponding abundance maps of 8 plasmonic components obtained by VCA processing. Hollow and solid sections are marked with white dashed lines.

Fig. 4D shows the 8 different plasmon components and their corresponding abundance maps obtained by applying VCA analysis to the EELS data of the hybrid AuAg nanotube. It should also be noted here that colors of the spectra obtained via VCA in Fig. 4D do not represent the colored regions in Fig. 4C. First thing to highlight in this figure is that the EEL spectrum and plasmon map of the first-order component located at 0.44 eV, which could not be observed after the zero-loss peak subtraction by Power Law. The components obtained by VCA revealed the second, third and fourth-order resonator modes (components II, III and IV). As shown experimentally and by BEM simulations for the completely hollow AuAg nanotubes, Fabry-Perot resonator modes are generated intensely inside the hollow nanotubes. The abundance map of the third-order mode (component III) clearly confirms such a behavior for the hybrid AuAg nanotube, where the same mode has a much higher intensity at the hollow part (on the right) compared to the solid part (on the left) along the same nanotube. Intriguingly, the distribution of resonator-like modes is not symmetrical for the hybrid AuAg nanotube, due to the plasmon hybridization between the hollow and solid parts of this 1D nanostructure. Yazdi et al. [35] reported such a plasmon hybridization in partially hollow, shorter AgAu nanorods. Here we show that symmetry breaking in resonant modes can manifested clearly for higher-order modes in longer and consequential hybrid 1D nanostructures. Liang et al. [60] also reported a similar asymmetric plasmon resonance distribution in asymmetric silver "nanocarrot" structures. Aside from these studies, plasmon mode coupling of higher-order modes was also observed in dimer/complex structures [29, 61, 62, 63, 64]. For instance, Schubert et al. [61] obtained a symmetry-break in AuAg nanowire dimers separated by 10 to 30 nm, where they reported a surface plasmon coupling between the second-order mode and third-order mode of two individual nanowires present in the asymmetric dimer, resulting in a bonding-antibonding mode pair. A component located at ∼2.3 eV, covering a quite wide range of energies between 1.6 and 3.1 eV is obtained (component VI) by VCA and its corresponding abundance map reveals that this component is associated with the LSPR mode of the hollow parts. Components related to the surface and bulk plasmon resonances of Ag are presented in components VII and VIII, respectively, along with their distribution indicating that they are only present in or surface of the solid parts.

Fig. 5 shows the BEM simulation results obtained from the hybrid AuAg nanotube presented in Fig. 4. During the BEM simulations, we used an AuAg nanotube with a length of 1240 nm and a diameter of 89 nm, with 10 nm thick walls (the same sizes as the experimentally investigated nanotube) with well-defined porous and hollow parts unlike the experimentally investigated one. The simulated hybrid AuAg nanotube model has a sequence of 50 nm hollow, 110 nm solid, 110 nm hollow, 340 nm solid, 375 nm hollow and 245 nm solid parts, where the hollow parts are composed of 60% Au and 40% Ag and the solid parts being pure Ag. Fig. 5A shows the simulated local EEL spectra obtained at the tip (in blue), at a quarter of the length (at ~310 nm, in green) and at the center (at 620 nm, in red) of the AuAg nanotube, where numerous plasmon peaks are observed.

By taking these peaks into account, we have obtained BEM simulated maps of 11 different plasmon modes (Fig. 5B) from the in-plane view. First 6 modes having plasmon resonances between 0.637 eV and 1.701 eV are Fabry-Perot resonator modes. It should be noted here that the experimental results presented in Fig. 4 reveal the presence of only 4 different resonator modes. Even though the number of the plasmon modes are not the same, the distribution of these plasmon resonances are quite similar for experimental and BEM simulated results. BEM simulated results confirmed the symmetry breaking for the Fabry-Perot resonator modes due to the plasmon hybridization between hollow and solid parts, and the experimentally observed plasmon intensity differences within the hybrid AuAg nanotube. A LSPR mode located at 1.891 eV is found to be mostly confined at the hollow tip of the hybrid nanotube, the distribution of which is quite similar to the experimentally observed plasmon mode located at ~1.6 eV. Hollow parts of the simulated nanotube have a LSPR mode located at 2.119 eV. As seen in the BEM simulated plasmon maps presented in Fig. 5B, two other LSPR modes generated from the Ag-rich regions are found to be located at 2.651 eV and 2.803 eV and the bulk plasmon resonance for the Ag parts are located at 3.81 eV.

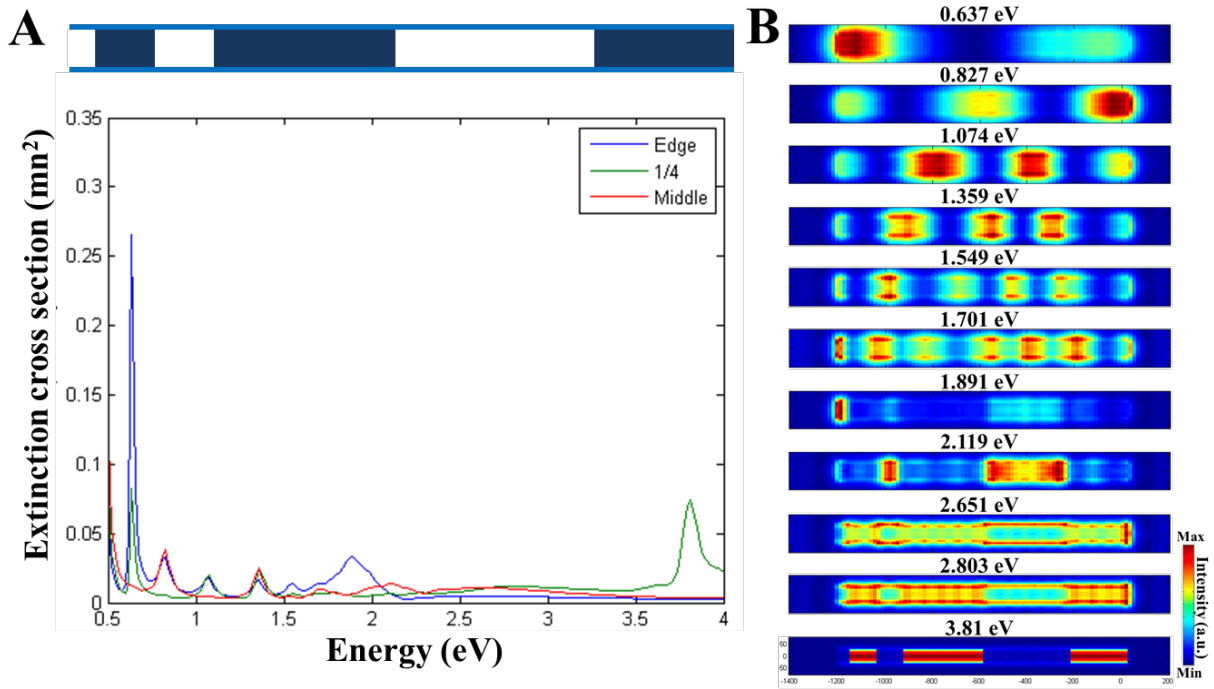

**Figure 5. BEM simulations of the hybrid AuAg nanotube. A.** Schematic drawing of the BEM simulated AuAg nanotube with a length of 1240 nm and a diameter of 89 nm, with 10 nm thick walls, along with the simulated local EEL spectra obtained at the tip (in blue), at a quarter of the length (at ∼ 310 nm, in green) and at the center (at 620 nm, in red) of the AuAg nanotube. **B.** BEM simulated plasmon maps of 11 different modes, located at 0.637 eV, 0.827 eV, 1.074 eV, 1.359 eV, 1.549 eV, 1.701 eV, 1.891 eV, 2.119 eV, 2.651 eV, 2.803 eV and 3.81 eV, from the in-plane view.

With the local plasmonic properties of these hybrid nanotubes and the recent advancement in the fabrication of hollow nanostructures via nanosecond laser [65] or localized electron beam irradiation, a new era on the controlled tailoring of codified plasmonic nanostructures with symmetric/asymmetric plasmonic properties can be initiated.

CONCLUSIONS

In this paper, the plasmonic properties of hollow metal 1D nanostructures are presented. Bimetallic AuAg nanotubes are synthesized via galvanic replacement process where solid Ag nanowires are used as a template resulting in the formation of nanotubes. We have studied samples consisting of completely hollow or hybrid AuAg nanotubes. The hybrid nanotube is a sequence formation of a solid Ag core with AuAg shell and hollow AuAg shell/wall parts. The full plasmonic properties of a completely hollow AuAg nanotube, presenting a length of 655 nm and a diameter of 84 nm with 10 nm thick walls are presented. The presence of several Fabry-Perot resonator modes and LSPR modes and their distribution are shown by VCA. Furthermore, it is shown that the plasmon modes, especially Fabry-Perot modes, are highly intense inside the nanotube. These experimental results are perfectly corroborated by BEM simulations obtained from an AuAg nanotube composed of 60 at.% Au and 40 at.% Ag with the same sizes as the experimentally investigated hollow AuAg nanotube.

The presence of multiple Fabry-Perot modes and LSPR modes are observed on a 1.24 µm long hybrid AuAg nanotube with a diameter of 89 nm. The LSPRs in the hybrid AuAg nanotube have been generated distinctively from the hollow and solid parts of the nanotube, which opens the way to control a broad range of plasmon resonances with one single nanostructure. The periodicity of the Fabry-Perot modes is disrupted in this hybrid AuAg nanotube due to the plasmon hybridization by the interaction of solid and hollow parts, which resulted in an asymmetrical plasmon distribution in a single 1D nanostructure. We believe that understanding the plasmon resonances of such novel nanostructures and the possibility of codifying the presence of hollow cavities in the nanotubes by applying laser or electron beam irradiation in localized areas open a new field in plasmonics for the accurate control of plasmon resonances at will.

ASSOCIATED CONTENT

**Supporting Information**. STEM-EDX analyses of the hybrid nanotubes. BEM simulations of pure Ag nanowires and Ag nanotubes in vacuum. BEM simulations of Ag nanotubes on a 15 nm thick $Si_3N_4$ substrate. The following files are available free of charge. Supporting Info_Hybrid NTs Plasmonics (PDF).

AUTHOR INFORMATION

**Corresponding Authors**

* Corresponding authors: azizgenc@iyte.edu.tr, arbiol@icrea.cat

**Author Contributions**

The manuscript was written through contributions of all authors. All authors have given approval to the final version of the manuscript.

ACKNOWLEDGMENT

ICN2 acknowledges funding from Generalitat de Catalunya 2021SGR00457. This study was supported by MCIN with funding from European Union NextGenerationEU (PRTR-C17.I1) and Generalitat de Catalunya. This research is part of the CSIC program for the Spanish Recovery, Transformation and Resilience Plan funded by the Recovery and Resilience Facility of the European Union, established by the Regulation (EU) 2020/2094. The authors thank support from the project NANOGEN (PID2020-116093RB-C43), funded by MCIN/ AEI/10.13039/501100011033/ and by "ERDF A way of making Europe", by the "European Union". ICN2 is supported by the Severo Ochoa program from Spanish MCIN / AEI (Grant No.: CEX2021-001214-S) and is funded by the CERCA Programme / Generalitat de Catalunya. Part of the present work has been performed in the framework of Universitat Autònoma de Barcelona Materials Science PhD program.The STEM-EELS/-EDX measurements were performed at the Laboratorio de Microscopias Avanzadas (LMA), Universidad de Zaragoza (Spain). R.A. acknowledges support from Spanish MCIN (PID2019-104739GB-100/AEI/10.13039/501100011033), Government of Aragon (project DGA E13-20R (FEDER, EU)) and from EU H2020 "ESTEEM3" (Grant number 823717). NGB and VP acknowledge financial support from the Spanish Ministerio de Ciencia, Innovación y Universidades (MCIU) (RTI2018-099965-B-I00, AEI/FEDER,UE). Authors acknowledge Prof. M. Duchamp for the VCA code.

# Supporting Information for

# Asymmetrical plasmon distribution in hybrid AuAg hollow/solid coded nanotubes

*Aziz Genç,[a,b,*] Javier Patarroyo,[a] Jordi Sancho-Parramon,[c] Raul Arenal,[d,e] Neus G. Bastús,[a] Victor Puntes,[a,f,g] Jordi Arbiol[a,g,*]*

[a] Catalan Institute of Nanoscience and Nanotechnology (ICN2), CSIC and BIST, Campus UAB, Bellaterra, 08193 Barcelona, Catalonia, Spain.

[b] Materials Science and Engineering Department, Izmir Institute of Technology, 35430, İzmir, Turkey.

[c] Rudjer Boskovic Institute, Zagreb, Croatia.

[d] ARAID Foundation, 50018 Zaragoza, Aragon, Spain.

[e] Laboratorio de Microscopias Avanzadas (LMA), Instituto de Nanociencia y Materiales de Aragon (INMA), Universidad de Zaragoza, 50018 Zaragoza, Spain.

[f] Vall d'Hebron Institut de Recerca (VHIR), 08035, Barcelona, Catalonia, Spain.

[g] ICREA, Pg. Lluís Companys 23, 08010 Barcelona, Catalonia, Spain.

 * Corresponding authors: azizgenc@iyte.edu.tr, arbiol@icrea.cat

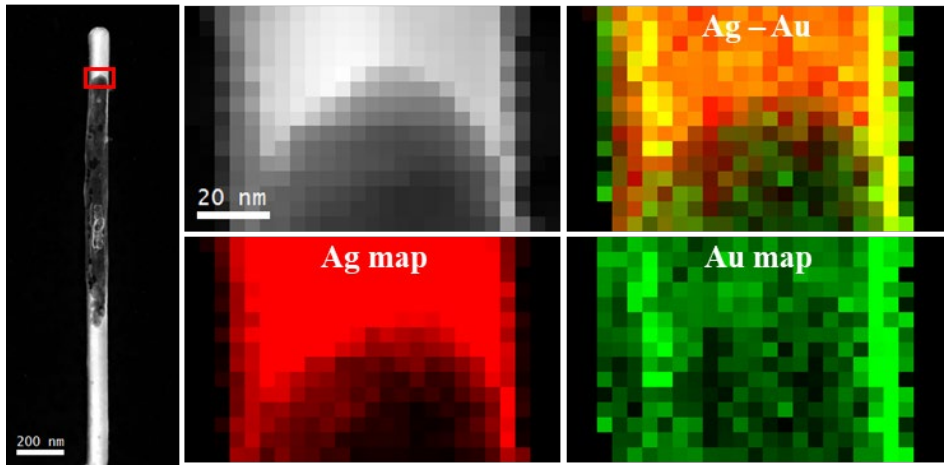

**Figure S1.** HAADF – STEM image of an individual hybrid AuAg nanotube and EDX elemental maps of Ag and Au over the area indicated with a red rectangle, along with their composite image.

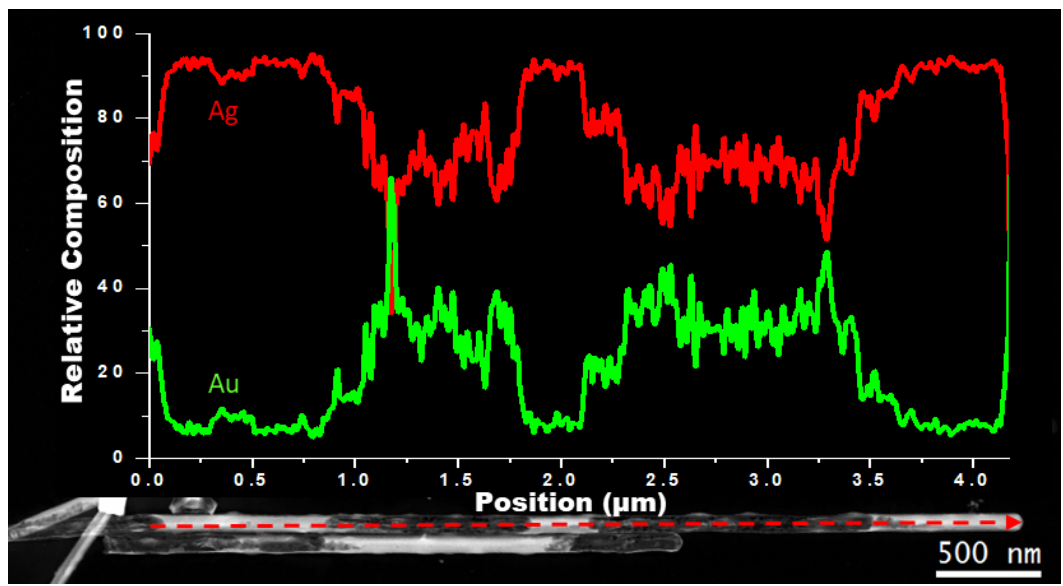

**Figure S2.** HAADF – STEM image of an individual hybrid AuAg nanotube and relative compositions of Ag (in red) and Au (in green) obtained by an EDX line scan along the red arrow.

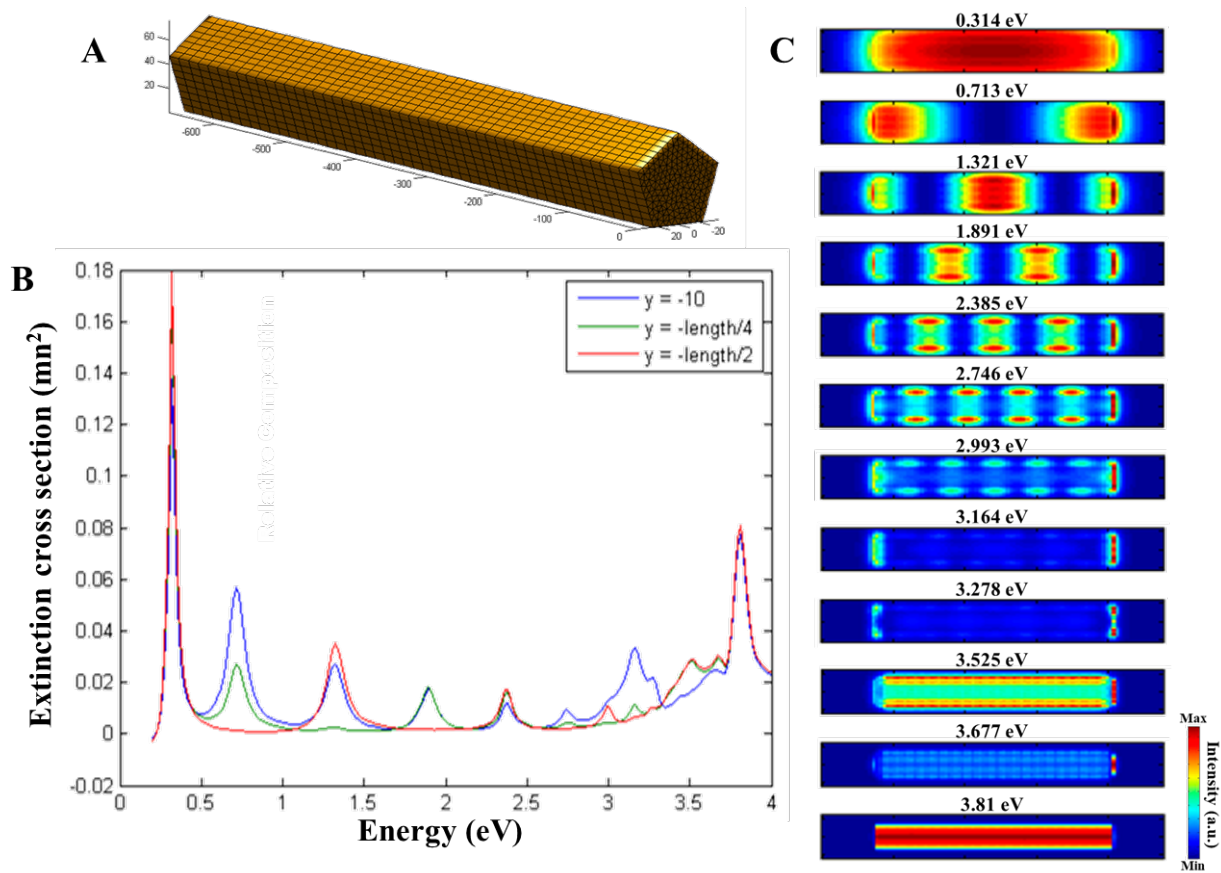

**Figure S3. BEM simulations of Ag nanowire.** A. Structural model of the BEM simulated Ag nanowire with a length of 665 nm and a diameter of 84 nm. B. Simulated local EEL spectra obtained at the tip, at a quarter of the length (at ~166 nm) and at the center (at 332.5 nm) of the Ag nanowire. C. BEM simulated plasmon maps of 12 different plasmon modes for the Ag nanowire. Note that the nanowire is standing in vacuum.

Since plasmon mapping of Ag nanowires/nanorods via EELS is studied extensively in the literature, we only conduct BEM simulations for Ag nanostructures. We simulate a Ag nanowire with the same dimensions as the completely hollow AuAg nanotube. Thanks to the applied BEM simulations, we also report the effects of hollow morphology for nanostructures standing in vacuum as well as the effects of presence of a silicon nitride substrate on the plasmonic properties of 1D Ag nanostructures.

Fig. S3A shows the structural model of the Ag nanowire that is used for the BEM simulations. The simulated nanowire has a pentagonal morphology with the same sizes of a 665 nm length and 84 nm diameter as the experimentally investigated (presented in Fig. 2 of the main text), completely hollow AuAg nanotube. It should be noted here that the Ag nanowire is standing in vacuum. Fig. S3B shows the BEM simulated EELS spectra obtained at the tip (near the edge, in blue), at the one quarter of the nanowire length (at ∼166 nm, in green) and at the center (at 332.5 nm, in red) of the nanowire. As seen in these EEL spectra, presence of several peaks are observed at the different locations of the Ag nanowire. At the first instance, the presence of at least 12 peaks can be observed. BEM simulated plasmon distribution maps of these peaks are presented in Fig. S3C. The lowest energy peak is located at 0.314 eV and it is present in all three locations, suggesting a presence of a dark plasmonic breathing mode for this simulated Ag nanowires. Second peak located at 0.713 eV is present at the tip with high intensity. The periodicity of the peak locations for some of the peaks listed above suggests the presence of Fabry-Perot resonator type modes, which is quite expected for an Ag nanowire. And, by looking at the distribution of the plasmon resonance peak located at 0.713 eV, it can be suggested that this is the first order resonator mode. A third peak located at 1.321 eV is present at the tip and at the center of the nanowire and is identified as the second order resonator mode. Plasmon distribution maps of the resonator modes up to sixth order are present in Fig. S3C (located at ∼1.89 eV, ∼2.38 eV, ∼2.74 eV and ∼2.99 eV, respectively), with some other localized surface plasmon resonance modes (located at ∼3.16 eV, ∼3.28 eV, ∼3.52 eV and ∼3.68 eV) and bulk plasmon mode of Ag at 3.81 eV.

BEM simulation results of an Ag nanotube are presented in Fig. S4, allowing us to understand the effects of hollow morphology on the plasmonic properties. Fig. S4A shows the structural model of the Ag nanotube that is used for the BEM simulations. The simulated nanotube also has a pentagonal morphology with the same sizes of a 665 nm length and 84 nm diameter with a 10 nm thick continuous wall. It is standing in vacuum. Fig. S4B shows the BEM simulated EELS spectra obtained at the tip (near the edge, in blue), at the one quarter of the nanotube length (at ~166 nm, in green) and at the center (at 332.5 nm, in red) of the nanotube. At the first instance, one can see that the breathing plasmon mode observed for the Ag nanowire is not present at the BEM simulated EEL spectra of the Ag nanotube. Moreover, it is clear that plasmon resonances shifted to lower energies compared to solid Ag nanowire due to plasmon hybridization between solid and cavity modes [1]. As seen in these EEL spectra, presence of several peaks are observed at the different locations of the Ag nanotube. It should be pointed out here that we did not take the instrumental broadening into account during these BEM simulations. Some of these peaks with very similar energy values would merge together in the experimental EELS measurements.

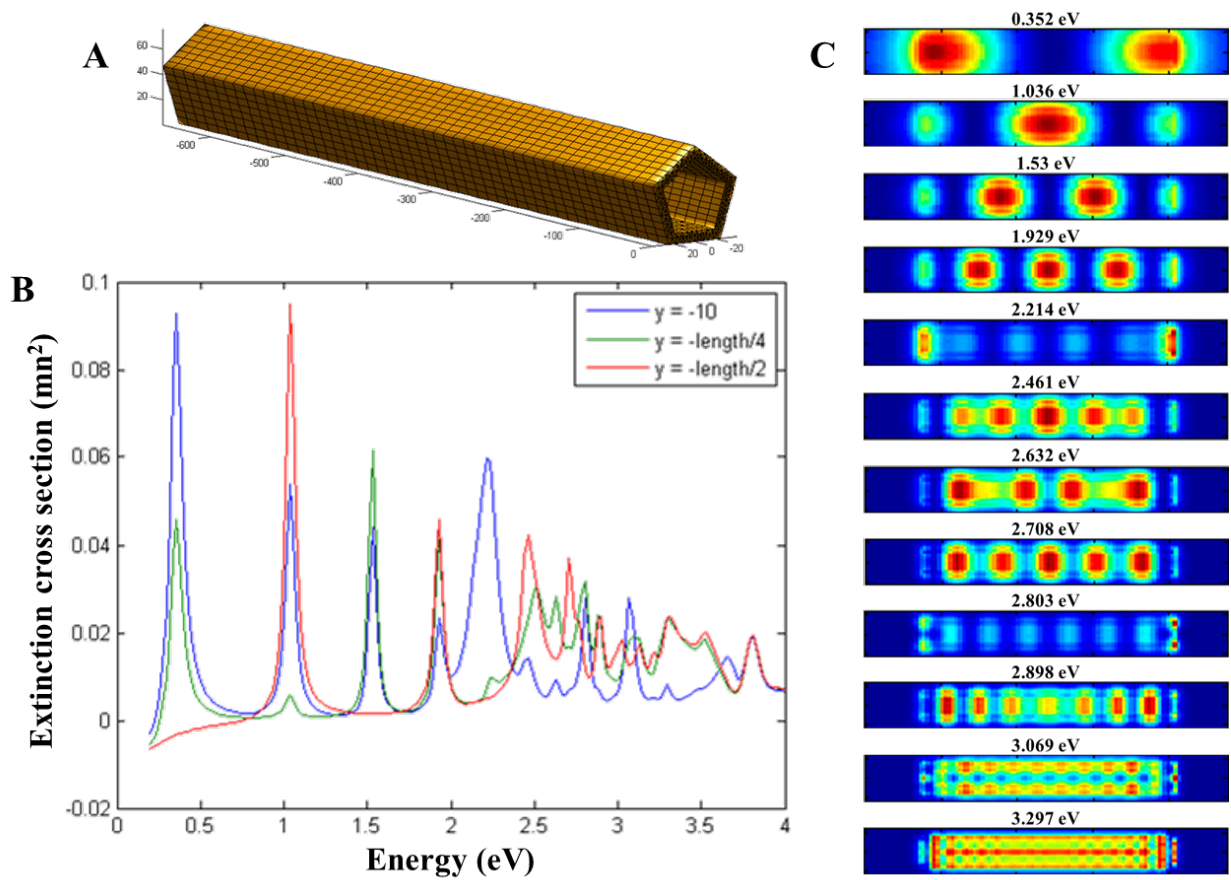

**Figure S4. BEM simulations of Ag nanotube.** A. Structural model of the BEM simulated Ag nanotube with a length of 665 nm and a diameter of 84 nm. B. Simulated local EEL spectra obtained at the tip, at a quarter of the length (at ∼ 166 nm) and at the center (at 332.5 nm) of the Ag nanotube. Note that the nanotube is standing in vacuum. C. BEM simulated plasmon maps of 12 different plasmon modes for the Ag nanotube.

BEM simulated plasmon maps of 12 different peaks located at 0.352 eV, 1.036 eV, 1.53 eV, 1.929 eV, 2.214 eV, 2.461 eV, 2.632 eV, 2.708 eV, 2.803 eV, 2.898 eV, 3.069 eV and 3.297 eV in Fig. S4B are shown in Fig. S4C. The mode located at 0.352 eV is the first order Fabry-Perot type resonator mode, which was located at 0.713 eV for the solid Ag nanowire. Such a shift reveal the effects of the plasmon hybridization in hollow nanostructures [2]. The distributions of second, third and fourth order Fabry-Perot modes located at 1.036 eV, 1.53 eV and 1.929 eV, respectively, are clearly seen in this BEM simulated plasmon maps. It should be noted here that these Fabry-Perot resonator modes are most efficiently excited at the inside of the nanotube. The LSPR mode located at 3.164 eV in the solid Ag nanowire shifted significantly to 2.214 eV in the hollow Ag nanotube, which is highly confined at the tips of the nanotube. The nature of the modes located at 2.461 eV, 2.708 eV and 2.898 eV is not clear in these plasmon maps, i.e. they are most probably high order Fabry-Perot modes, yet they may be LSPR modes as well. In any case, it is clear that they have intense resonances both at the inner and outer parts of the nanotube. Similarly, a LSPR mode located at 2.632 eV have high intensities at the inner and outer parts of the nanotube. Another LSPR mode located at 2.803 eV is mostly confined at the tips with some contributions from the other parts of the nanotube. The plasmon maps of the modes located at 3.069 eV and 3.297 eV reveals that these modes have multipolar contributions.

After revealing the differences between a solid and hollow 1D nanostructure standing in vacuum, we continue with the addition of a substrate in order to discuss its effects on the plasmonic properties of hollow 1D nanostructures. As reported in the literature, plasmon modes of the nanostructures interact with the sample resulting in the formation of distal and proximal modes [3, 4]. Fig. S5A shows the structural model of the Ag nanotube (655 nm long, 84 nm wide with 10 nm thick walls) standing on a 15 nm thick $Si_3N_4$ substrate, which is used during BEM simulations of the Ag nanotubes. Fig. S5B shows the BEM simulated EELS spectra obtained at the tip (near the edge, in blue), at the one quarter of the nanotube length (at ~166 nm, in green) and at the center (at 332.5 nm, in red) of the nanotube. It is seen that all the peaks shifted to lower energies except the lowest energy peak corresponding to the first order Fabry-Perot resonator mode located at ~0.52 eV, which is located at 0.352 eV for the Ag nanotube standing in vacuum. As there is no study about the plasmonic properties of hollow 1D nanostructures, we are not sure whether this is caused by a numerical accuracy or it is typical for such nanostructures. The presence of many different peaks obtained at different parts of the Ag nanotube standing on a $Si_3N_4$ substrate is shown in this figure. Since there are more than 10 peaks located between 2 eV and 3.8 eV, we do not mention all the peaks with their energy values but one can easily distinguish several sharp peaks located at ~0.52 eV, ~1.0 eV, ~1.4 eV, ~1.7 eV eV, ~2.0 eV and ~2.4 eV among many others.

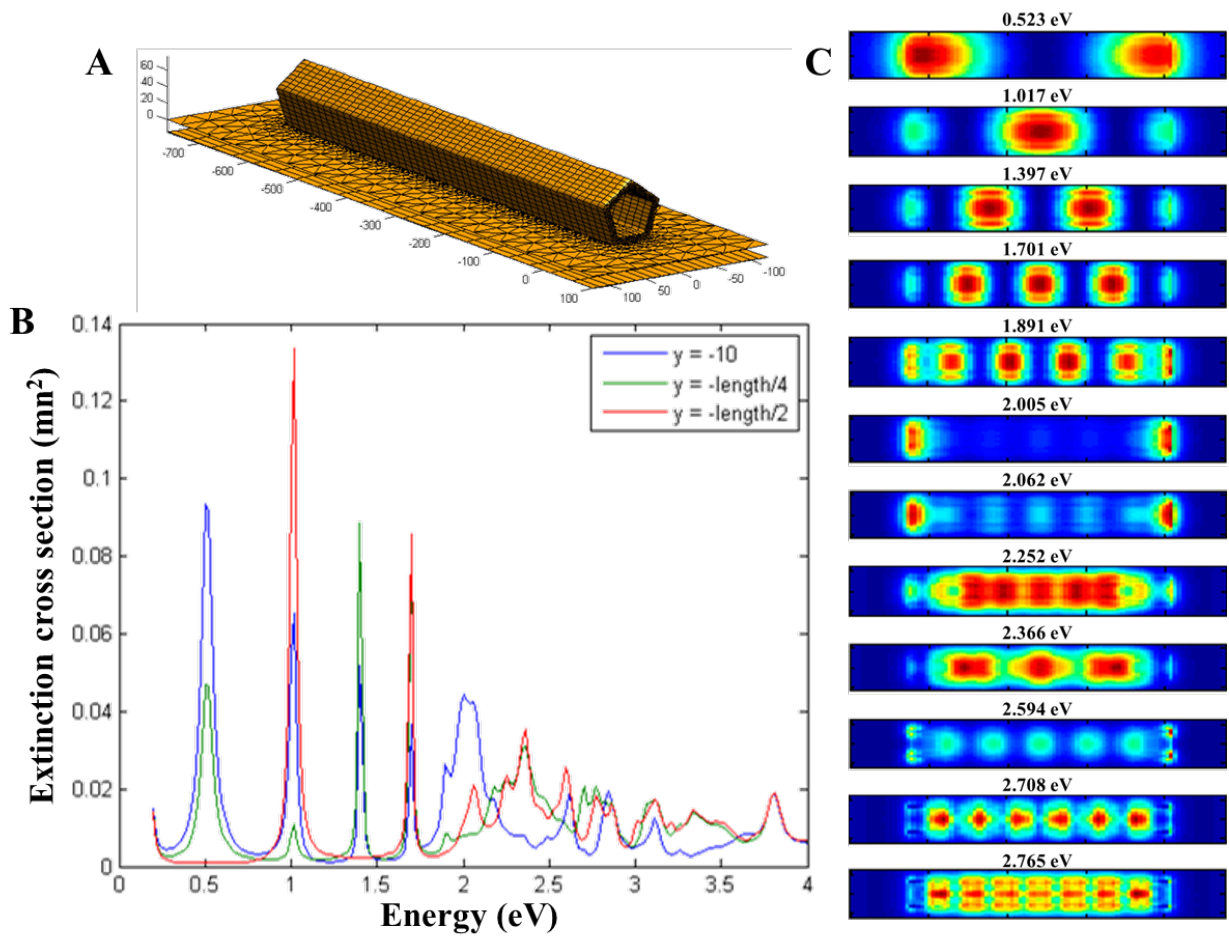

**Figure S5. BEM simulations of an Ag nanotube on 15 nm thick Si₃N₄ substrate.** A. Structural model of the BEM simulated Ag nanotube with a length of 665 nm and a diameter of 84 nm, with 10 nm thick walls, standing on a 15 nm thick Si₃N₄ substrate. B. Simulated local EEL spectra obtained at the tip, at a quarter of the length (at ~166 nm) and at the center (at 332.5 nm) of the Ag nanotube. C. BEM simulated plasmon maps of 12 different plasmon modes for an Ag nanotube standing of a Si₃N₄ substrate.

BEM simulated plasmon maps of 12 different peaks located at 0.523 eV, 1.017 eV, 1.397 eV, 1.701 eV, 1.891 eV, 2.005 eV, 2.062 eV, 2.252 eV, 2.366 eV, 2.594 eV, 2.708 eV and 2.765 eV in Fig. S5B are shown in Fig. S5C. As mentioned before, the peaks shifted to lower energies in general except the one located at 0.523 eV. Its BEM simulated map shows that this is the first order Fabry-Perot resonator mode and it is mostly excited inside the nanotube. Second, third, fourth and fifth order Fabry-Perot modes, which are located at 1.017 eV, 1.397 eV, 1.701 eV and 1.891 eV, respectively, reveals the similar distribution of plasmon resonances, i.e. they are mostly intense at the inner part of the Ag nanotube along with intense presence at the outer parts. Two LSPR modes that are located at ~2 eV and 2.062 eV are highly confined at the tips of the nanotube. As mentioned before, after taking the instrumental broadening into account, which is about 130 meV for the experimentally obtained EELS maps, these two peaks with an energy difference of 57 meV would merge together. BEM simulated plasmon maps of some other LSPR modes located at various energies are shown in this figure. Similar to the discussions about BEM simulated plasmonic properties of cuboid AuAg nanostructures, we have used pure Ag as a playground in order to discuss the plasmonic property differences between solid and hollow 1D nanostructures and the effects of substrate presence on the plasmonic properties of the hollow 1D nanostructures. So far, we have shown that both solid (Ag nanowires) and hollow (Ag nanotubes) 1D nanostructures contain multiple Fabry-Perot type resonator modes and LSPR modes. The presence of a dark plasmonic breathing mode for the Ag nanowire is observed during the BEM simulations and we need to have a better understanding about this mode. In addition, we have observed that the energy of Fabry-Perot and LSPR modes shifted to lower energies for the hollow nanotube compared to those of the solid nanowires, due to plasmon hybridization between solid and cavity modes. It has been revealed that most of the plasmon modes (both Fabry-Perot type and LSPR) are excited most intensely from the inner parts of the hollow nanostructure. Implementation of a 15 nm thick

Si3N4 substrate caused a furhter shift of the plasmon resonances of the Ag nanotubes to lower energies, except the first order Fabry-Perot mode which shifted to higher energies.

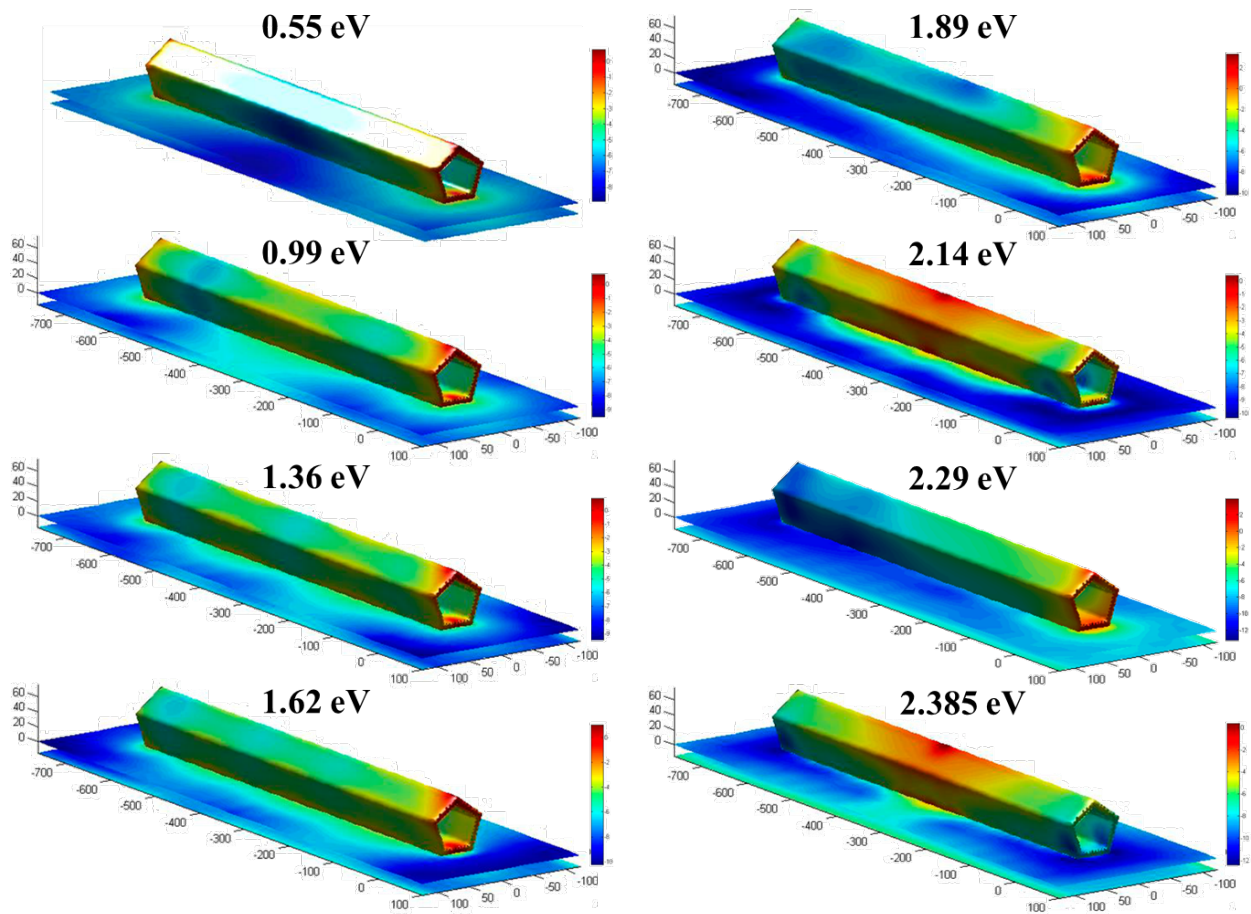

**Figure S6.** 3D BEM simulations of the hollow AuAg nanotube on 15 nm thick $Si_3N_4$ substrate.